# A complete data processing workflow for CryoET and subtomogram averaging

Muyuan Chen[1], James M. Bell[1,2], Xiaodong Shi[1,3], Stella Y. Sun[4], Zhao Wang[1,5], Steven J. Ludtke[1,5]

[1] Verna Marrs and McLean Department of Biochemistry and Molecular Biology, Baylor College of Medicine, Houston, Texas, USA

[2] Quantitative and Computational Biosciences Graduate Program, Baylor College of Medicine, Houston, Texas, USA

[3] Jiangsu Province Key Laboratory of Anesthesiology and Jiangsu Province Key Laboratory of Anesthesia and Analgesia Application, Xuzhou Medical University, Xuzhou, Jiangsu, China

[4] Department of Bioengineering, Stanford University, Stanford, California, USA

[5] CryoEM Core at Baylor College of Medicine, Houston, Texas 77030, USA.

## Abstract

Electron cryotomography (CryoET) is currently the only method capable of visualizing cells in 3D at nanometer resolutions. While modern instruments produce massive amounts of tomography data containing extremely rich structural information, the data processing is very labor intensive and results are often limited by the skills of the personnel rather than the data. We present an integrated workflow that covers the entire tomography data processing pipeline, from automated tilt series alignment to subnanometer resolution subtomogram averaging. This workflow greatly reduces human effort and increases throughput, and is capable of determining protein structures at state-of-the-art resolutions for both purified macromolecules and cells.

## Introduction

Electron cryomicroscopy (CryoEM) is rapidly becoming the standard tool for near atomic resolution structure determination of purified biomolecules over 50 kDa. However, for studies of molecules within cells or purified molecules that exhibit significant conformational variability, electron cryotomography (CryoET) is the preferred method[1]. In these experiments, the specimen is tilted within the microscope providing 3D information about each molecule and permitting overlapping densities, such as those found in the crowded cellular cytosol, to be isolated.

While recent microscope and detector advances have greatly boosted the throughput of CryoET data collection, substantial human effort and computational resources are still required to process recorded imaging data. Especially in cellular tomography projects, data processing has become a major bottleneck in studying high-resolution protein structures.

To expedite cellular tomography data processing, we present a complete tomography workflow as part of the EMAN2 environment that performs all steps within the standard CryoET data processing pipeline, from the raw tilt series alignment through subtomogram averaging. While many of these tools are based on decades of development by many groups[2–10], numerous innovations have been introduced to reduce human intervention and improve the resolution of the final average. These include a fully automated tilt-series alignment method not requiring fiducials, rapid 3D reconstruction using direct Fourier methods with tiling, an optimization-based strategy for per-particle-per-tilt CTF correction, robust initial model generation, and per-particle-per-tilt orientation refinement (Fig. 1a). In addition to algorithm development, this protocol also includes a user-friendly graphical interface and a specially designed book-keeping system for cellular tomography that allows users to study multiple features within one dataset, and to keep track of particles to correlate structural findings with the location of proteins in the cellular environment.

Our integrated pipeline significantly increases the throughput of CryoET data processing and is capable of achieving the state-of-the-art subtomogram averaging results on both purified and *in situ* samples. We demonstrate subnanometer resolution from previously published datasets[11], and cellular tomography of whole *e. coli* over-expressing a double-layer spanning membrane protein at 14 Å resolution.

## Results

### Automated tilt series alignment and tomogram reconstruction

The first stage of the tomogram processing workflow is tilt-series alignment. Our method uses an iterative landmark-based approach with progressive downsampling and outlier elimination (Fig. 1b). It works well on a wide range of tomograms with or without fiducials and without any human intervention.

The method begins with a coarse, cross-correlation based alignment of a downsampled tilt series, and a rough estimate of the orientation of the tilt axis via common line methods. The input tilt series are downsampled to 512x512 pixels irrespective of their original size or sampling. Based on the coarse alignment, an initial tomogram is generated, despite the likelihood of significant alignment errors, and 3D landmarks are selected from the resulting volume to use in the next stage of alignment. These landmarks are simply the N darkest voxels in the downsampled map, with a minimum distance constraint (Fig. 2b). When fiducials are present in the data, they will tend to be selected as landmarks, as long as they are sufficiently well-separated, but they are explicitly identified as such.

The next step is the iterative alignment. This includes two steps: refinement of landmark coordinates and optimization of the tilt images transforms. First, 3D coordinates of the selected landmarks are projected back to the tilt series, and corresponding 2D patches are extracted from the tilt images. Local subtomograms are reconstructed from the sub-tilt series of each landmark, to provide a more accurate center of mass for each. Then, 2D patches are re-extracted from the tilt images using the refined landmark positions, and calculate the translational alignment that centers each landmark in each extracted 2D patch. A global optimization algorithm is used to adjust the 3D tilt transforms such that center of all landmarks in 2D patches match the projected coordinates of the landmarks to the greatest possible extent. With these improved alignment parameters, a new tomogram is generated with better alignment which is used during the next round of reprojection and alignment. To improve convergence and increase the speed of alignment, the process begins with highly downsampled images and gradually increases sampling as alignment error decreases, finishing with the unbinned tilt series in the final iteration. A specified fraction of the worst matching landmarks is normally excluded in each iteration, and this is critical to obtaining a self-consistent consensus alignment.

In most tomograms it is convenient for slice-wise visualization and annotation if the X-Y plane is parallel to the ice surface. It is assumed that on average the landmarks will be coplanar with the ice, and thus this plane is rotated to become flat, based on principal component analysis of the landmark coordinates (Fig. 2e).

Tomogram reconstruction is performed using direct Fourier inversion rather than real-space methods such as filtered back projection[2] or SIRT[12]. Fourier methods have gradually become the standard in single particle reconstruction, but due to the size of tomographic volumes and concerns about edge effects and image anisotropy, most tomography software still uses real space methods[9,13]. We have adopted a Fourier reconstruction approach using overlapping tiles, which significantly reduces edge effects and memory requirements, while still remaining computationally efficient. For convenience, the tile size is defined by the reconstruction thickness, such that each tile is a cube. The overlapping tiles are individually reconstructed, then averaged together using a weighted average with a Gaussian falloff (Fig. S1).

Although the tilt series alignment is performed using the original full-sized images, the reconstructed tomograms are normally downsampled to provide sufficient resolution for visual inspection, annotation, and particle selection, while dramatically improving interactivity and decreasing system requirements. For subtomogram averaging, the particle data is automatically extracted from the original tilt images to take advantage of the full sampling of the original data. The combined alignment and reconstruction algorithm is quite rapid, typically requiring only ~10 minutes on a 12-core workstation for full-resolution alignment of a 60 image 4k x 4k tilt series with a 2k x 2k x 512 downsampled reconstruction (Table S1). Since this is comparable to the time required for tilt series acquisition, it would be feasible to include this as an automated process during data collection.

As a test of this process, we reconstructed a cellular tomogram of *e. coli* over-expressing Tolc-AcrAB (Fig. 2a, Supplementary video 1)[14]. The improved alignment after this iterative process can be observed by comparing the reconstructions of fiducials before and after the iterative process. Internal cellular features are also clearly visible in the reconstruction. In fiducialless reconstructions, the program usually chooses small pieces of ice contamination or other high-density objects as landmarks (Fig. 2d). For fiducialless apoferritin data (EMPIAR-10171)[15], the program produced high quality reconstructions where individual proteins were clearly visible (Fig. 2c, Supplementary video 2).

**Multiple methods for particle localization**

Earlier versions of EMAN2 included a graphical program for manually selecting 3D particles using orthogonal slices[16]. In the latest version, this particle picking interface has been reworked, enabling users to simultaneously select and visualize particles of multiple types and different sizes within each tomogram (Fig. 3a,c). Each type is then extracted into a separate stack of 3D particles and accompanying 2D subtilt series, with the original location metadata retained for later per-particle processing.

In addition to the manual 3D picking interface, two semi-automatic tools are provided for annotation and selection. For purified macromolecules imaged by tomography, a template matching algorithm can be used to rapidly locate particles. For more complex tomograms, our convolutional-neural-network-based tomogram annotation tool can be used to identify features[17], followed by a second stage which converts annotations into subtomogram coordinates. For globular particles like ribosomes, the program locates and centers isolated annotations. For continuous structures like microtubules and protein arrays on membranes, the program randomly samples coordinates within the set of annotated voxels, with a specified minimum distance between boxes. The parameters of these semi-automatic tasks can then be tuned by visualizing results in the manual particle picking tool.

**Per-particle-per-tilt CTF correction**

Accurate CTF measurement and correction is critical for obtaining high-resolution structures through subtomogram averaging. The most commonly used method in tomographic CTF correction is the simple tiled CTF correction of rectangular strips within each tilt series[18]. This method is effective in getting past the first CTF zero-crossing when working with thin layers of purified macromolecules; however, when working with cellular data or other thicker specimens, the error in defocus due to the Z position of the particle within the ice becomes significant and requires more accurate correction on a per-particle per-tilt basis.

In the new CTF estimation strategy, the entire tilt image is used to determine its central defocus, by splitting the image into tiles and summing the information from the entire image to estimate the defocus. To do this, we find the defocus value $d$ that maximizes $\sum_i S_i(p_i, d + x_i \sin(\theta))$, where $x_i$ is the $x$ position of the $i$th tile ($y$ is the tilt axis), $\theta$ is the tilt angle, and $S_i(p, \Delta z)$ is the score function represented by the normalized dot product between a theoretical CTF curve with defocus $\Delta z$ and the coherent, background subtracted power spectrum, $p$, of the $i$th strip of tiles parallel to the tilt axis. With this approach, the information in the full tilt image is used to estimate one scalar value and achieve more robust defocus estimation in low SNR conditions.

At high tilt, the SNR in an individual image is typically so low that even using all information in the image is not sufficient to provide an unambiguous defocus estimation. Thus, for the higher tilts, only defocus values within three standard deviations around the mean defocus of the low tilt images are considered. With this additional constraint, reasonably accurate defocus values can be determined at high tilt.

After CTF determination, fully sampled CTF corrected subtomograms are generated directly from the raw tilt series. Since we have the alignment parameters for each micrograph in the tilt series and the coordinates of particles in the tomogram, we can extract per-particle tilt series, which we henceforth refer to as a set of subtilts from 2D micrographs. The center of each subtilt is determined by projecting the 3D coordinates of the particle using the transform of the micrograph calculated from tilt series alignment, so each subtilt series can be reconstructed to an unbinned 3D particle using the corresponding tilt image transforms. From these defocus values at the center of each tilt, the defocus of each tilt for each particle can be determined from the 3D location of the particle and the tilt-series geometry (Fig. S2). After subtilt images are extracted from the tilt series, we flip the phase of each subtilt according to its determined defocus before reconstructing the subtilt into CTF corrected 3D subtomograms.

**Initial model generation via stochastic gradient descent**

In many cellular tomography projects, the identities of extracted particles are unknown before subtomogram averaging. While it is possible to use catalogs of potential candidate structures and exhaustively compare particles to each of these for purposes of identification[19], there are many shortcomings to this approach, including the need for a complete catalog, the problem of model bias, and the difficulty of handling complexes. An unbiased approach would be to classify particles de-novo and generate independent initial models for each class from the raw particles. Our previous subtomogram averaging method offered several different strategies for handling this issue, as the failure rate was substantial. We have now developed a stochastic gradient descent (SGD) based initial model generation protocol[20], which produces reliable initial models even from cell-derived particles.

SGD is an optimization technique widely used in the training of machine learning models, which offers advantages in both speed and avoidance of local minima. We begin with an effectively randomized map, produced by averaging a random subset of particles in random orientations, lowpass filtered to 100 Å. In each iteration, a batch of randomly selected particles are aligned to the reference map, and a new map is generated. This new map is used to update the reference using an adjustable learning rate. To avoid overfitting, the reference is filtered to a user-specified resolution (usually 30-50 Å) after each update. The alignment, average and map update steps are repeated until the reference map converges to a consistent initial model. As only a low-resolution initial model is needed, it is not critical that all particles be used. The program can typically produce good initial models within 1 hour on a typical workstation (Table S1).

In testing, this method has performed well for structures with very distinct shapes from a variety of sources. This includes globular structures like ribosomes, linear structures such as microtubules, and even double-membrane spanning proteins (Fig. 3b,d).

**Subtomogram alignment and averaging**

There are two stages in producing a final high-resolution subtomogram average: traditional subtomogram alignment and averaging[5,16] and per-particle-per-tilt refinement (Fig. 1c). The initial stage makes use of our existing subtomogram alignment and averaging algorithms which automatically detect and compensate for the missing wedge[6]. The alignment algorithm employs an extremely efficient hierarchical method, which scales well with particle dimensions. The overall refinement process follows "gold-standard" procedures similar to single particle analysis[21], in which even and odd numbered particles are processed completely independently with unique, phase-randomized starting models, with a Fourier shell correlation (FSC) used to filter the even and odd maps, assess resolution, and measure iteration-to-iteration convergence.

In the second stage, rather than working with subtomograms, we work instead with subtilt series. When full frame tilt series are aligned, we assume that each tilt is a projection of a single rigid body volume. With beam-induced motion, charging and radiation damage affects the assumption that the specimen remains globally rigid across a 1-µm span with the largest acceptable motion <10 Å is an extremely stringent requirement. Local deviations are common and can produce significant misalignments of individual objects in individual tilts. To compensate for this resolution-limiting effect, we have developed a strategy for refinement on a per-particle-per-tilt basis, where the alignment and quality assessment of each tilt of each particle are individually refined. Effectively this is a hybridization of subtomogram averaging approaches with traditional single particle analysis. Some of these techniques are similar to those recently implemented in EMClarity[8].

Our subtilt refinement procedure starts from an existing 3D subtomogram refinement, preferably with a resolution of 25 Å or better. Subtilt series for each particle were already extracted as part of the CTF correction process above. The iterative refinement process is a straightforward orientation optimization for each tilt image of each particle. All 5 orientation parameters are refined independently per-particle-per-tilt. It is quite common for some images in a tilt series to be bad, either due to excessive motion or charging. To compensate for this, the quality of each tilt for each particle is assessed, and weighted correspondingly, with the very worst excluded entirely. All of the realigned particles are used to compute a new weighted average 3D map, which is then used for the next iteration of the refinement.

The subtilt refinement protocol significantly improves map quality and resolution for purified samples in thin ice, where relatively little density is present above and below each particle. In the EMPIAR-10064 dataset (purified ribosomes)[11], without subtilt refinement, subtomogram averaging achieved 13 Å "gold-standard" resolution (FSC>0.143) using 3000 particles from 4 tomograms. With subtilt refinement, the resolution improved dramatically, to 8.5 Å (Fig. 4a-d,

Supplementary video 3). In the averaged map, the pitch of RNA helices is clearly visible and long alpha-helices are separated.

We did not expect subtilt refinement to work well in a cellular context, due to the presence of so much confounding cellular mass present in each subtilt image. Surprisingly, we found that an *in situ* dataset of the double-membrane spanning TolC-AcrAB complex in *e. coli*[14], reached 19Å in initial averaging, which improved to 14 Å resolution after subtilt alignment (Fig. 4e, Supplementary video 4). We do not yet have sufficient test cases to set expectations for how well subtilt refinement will work in any given cellular system, but based on our preliminary studies, it may provide a significant improvement in a wide range of experimental situations.

**Discussion**

The entire protocol outlined above has been integrated into the graphical workflow in EMAN2.22 (e2projectmanager.py). This presents the process as a sequence of steps (Fig. 1), and an online tutorial can be found at http://eman2.org/Tutorials. Graphical tools are also provided for evaluating tomogram reconstructions and subtomogram refinements, which are useful for managing projects involving a large amount of data. Unlike single particle analysis where it is possible to transition data from other tools into EMAN2 at virtually any stage of processing, the stringent requirements for all of the metadata generated at each stage of processing make it challenging to, for example, import a reconstructed tomogram from other software, then proceed. While some tools will be usable on imported data, such as the Deep Learning based annotation and simple subtomogram alignment and averaging, the new approaches, such as subtilt refinement, are simply not possible unless the complete EMAN2 pipeline is followed.

With per particle CTF correction and subtilt refinement, it is now relatively straightforward to achieve ~10Å resolution using 1000 - 2000 particles from a few good tilt series. This method can also be used with phase-plate data, though the difficulty of collecting Volta phase plate tilt series and determining per-tilt CTF parameters with continuously varying phase shift is significant. While we do optimize both the defocus and phase shift, particularly at high tilt, there is insufficient information available for simultaneous determination of both parameters. Our suggested approach is to target 0.5 - 1 μm underfocus with such tilt series, to put the first zero in a range where correcting beyond the second zero is not necessary to achieve slightly better than 10Å resolution. In this way locating the first zero accurately is sufficient for subnanometer resolution.

One difficulty in subtomogram averaging *in situ* is masking and filtration of the averaged map after each iteration of refinement. In the cellular environment, proteins of interest are often surrounded by other strong densities and masking can have a strong impact on the final achieved resolution. To address this issue, we introduce the option of masking the averaged map with a large soft mask and filter it using the local resolution determined from even and odd sub-maps. This allows us to keep high-resolution information of the protein of interest for the next round of refinement and reduces misalignment caused by other densities surrounding the protein.

The algorithmic improvements we have discussed make it possible to perform data-driven cellular-structural biology research with CryoET. Researchers can take tomograms of cells or purified organelles, manually select a few features of unknown identity, and automatically annotate similar features in the whole dataset. Reliable, de novo initial models of the features of interest can be generated from raw particles without prior knowledge of the proteins. With per particle CTF correction and subtilt refinement, averaged maps at 10-15 Å resolutions can be achieved in a matter of days (Table S1) with a few thousand subtomogram particles, so one can make reasonable hypotheses of the identity and composition of the proteins based solely on their structural features, and validate these hypotheses with biochemical experiments. Furthermore, the position and orientation of each protein particle can be mapped back to the tomogram to study the organization of proteins in cells (Fig. 4f).

## Acknowledgments

This work was partially supported by NIH grants R01GM080139, P01GM121203, Welch Foundation (Q-1967-20180324), BCM BMB department seed funds, and a Houston Area Molecular Biophysics Program (HAMBP) training grant from the Keck Center of the Gulf Coast Consortium (GCC, T32 GM008280-30). We also would like to thank early users for testing the workflow and providing valuable feedbacks.

## Author contributions

M.C., J.M.B and S.J.L. designed and implemented the protocol. X.S., Z.W. and S.Y.S. provided test datasets. M.C., J.M.B. and S.Y.S. tested and refined the protocol. M.C., J.M.B. and S.J.L. wrote the manuscript.

## Data availability

The subtomogram averages are depositted to EMDatabank. EMD-0529: averaged structure of purified ribosome. EMD-0530: averaged structure of AcrAB-TolC from cellular tomogram.

## Competing financial interests
The authors declare no competing financial interests.

**Figures**

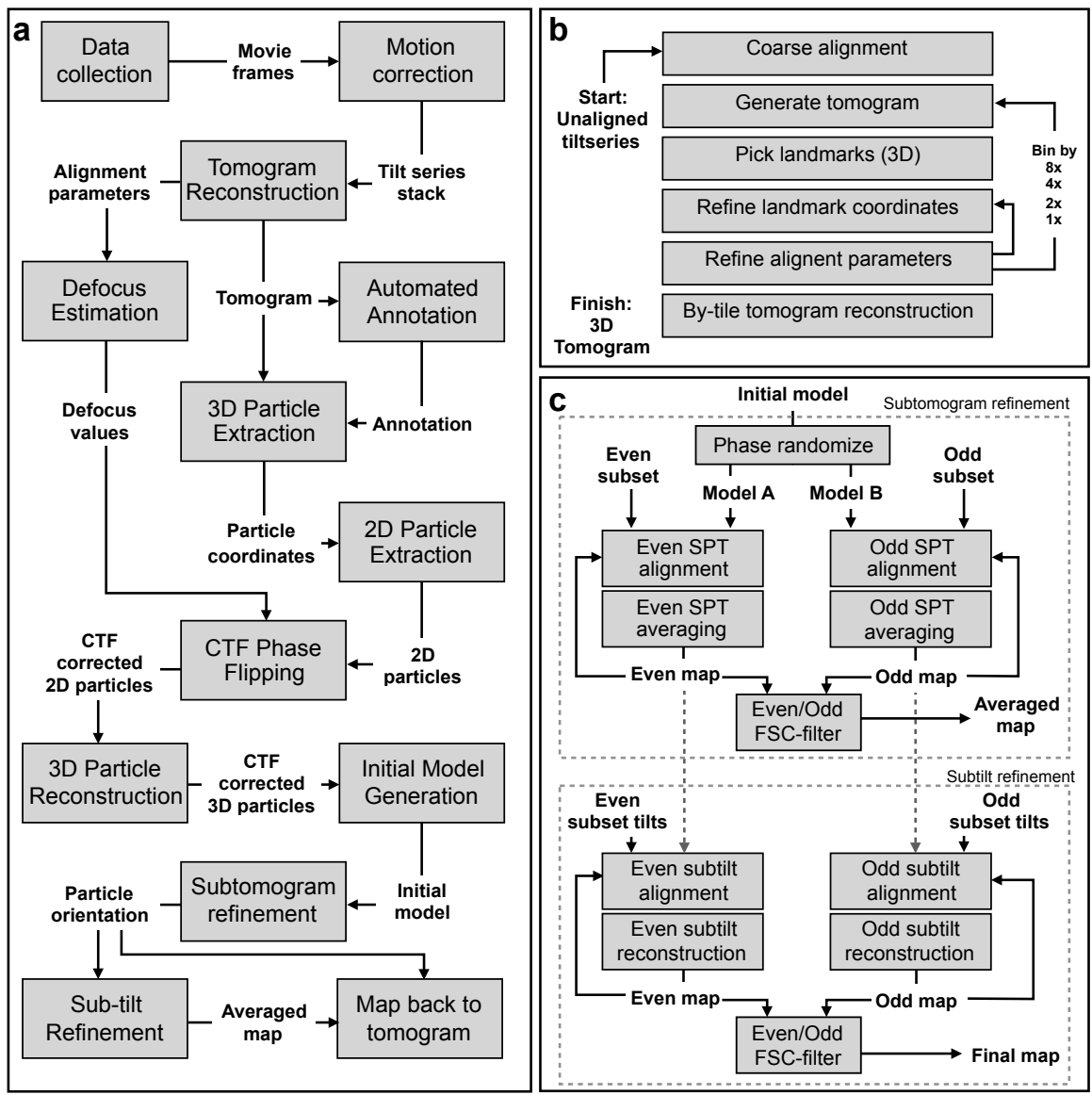

Fig 1. Workflow. (a) Main workflow diagram. (b) Workflow of tomogram reconstruction. (c) Workflow of subtomogram refinement and subtilt refinement.

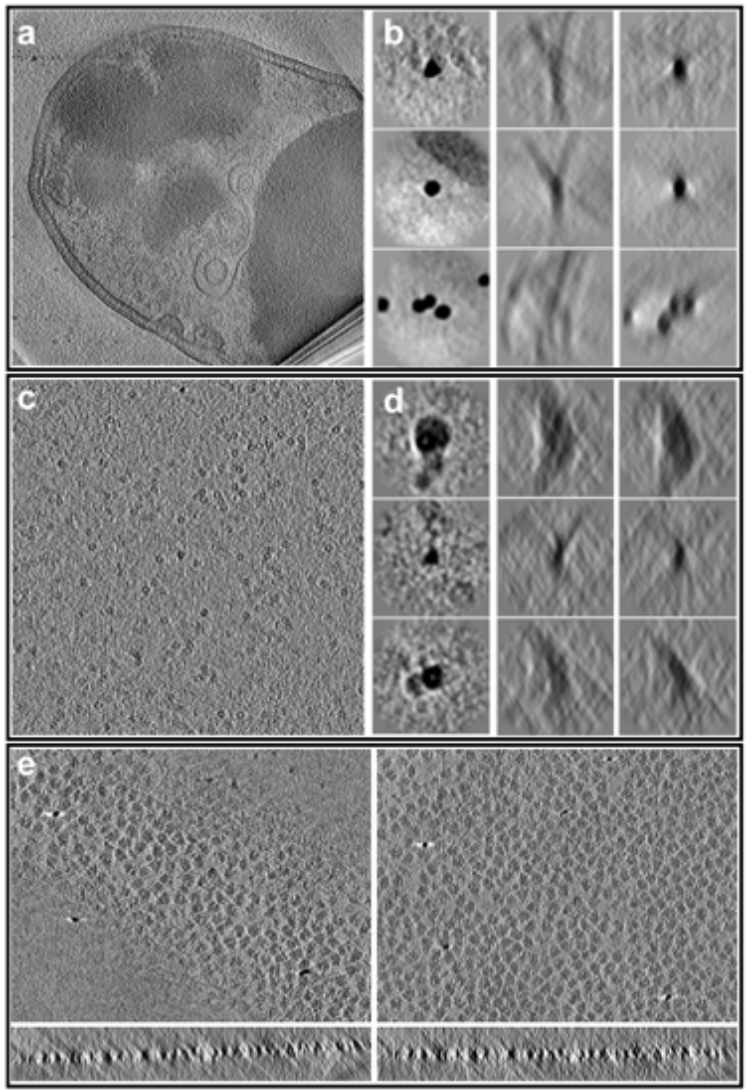

Fig 2. Results of iterative tomogram alignment and reconstruction (a) Cellular tomogram of e. coli with gold fiducials. (b) Selected landmark projections from (a) (left) x-y plane; (mid) x-z plane after the first iteration of the iterative alignment; (right) x-z plane after iterative alignment. (c) Tomogram of purified apoferritin without fiducials (EMPIAR-10171). (d) Selected landmark projections from (c). (e) Automatic specimen plane alignment. Left: (top) x-y slice (bottom) x-z slice, both before alignment; right: after alignment.

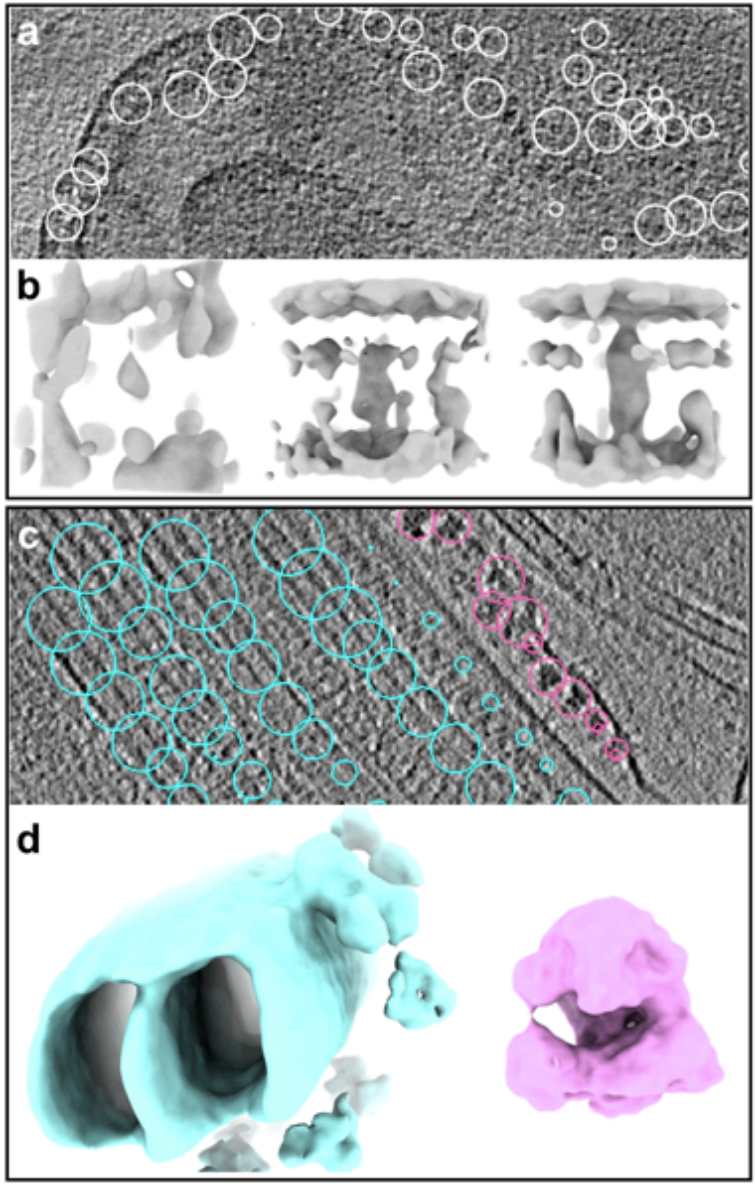

Fig 3. Particle extraction and initial model generation. (a) Slice view of a e. coli tomogram with particles of Tolc-AcrAB pump selected. (b) Initial model generation from Tolc-AcrAB pump particles. From the left to right are density maps of the initial seed, after 5 iterations with c1 symmetry, and after 5 iterations with c3 symmetry. (c) A tomogram slice view of the flagellum of an anucleated *Trypanosoma brucei* cell, with cyan circles selecting microtubule doublets, and pink circle selecting ribosomes. (d) Initial model generated from microtubule (left) doubles and ribosomes (right).

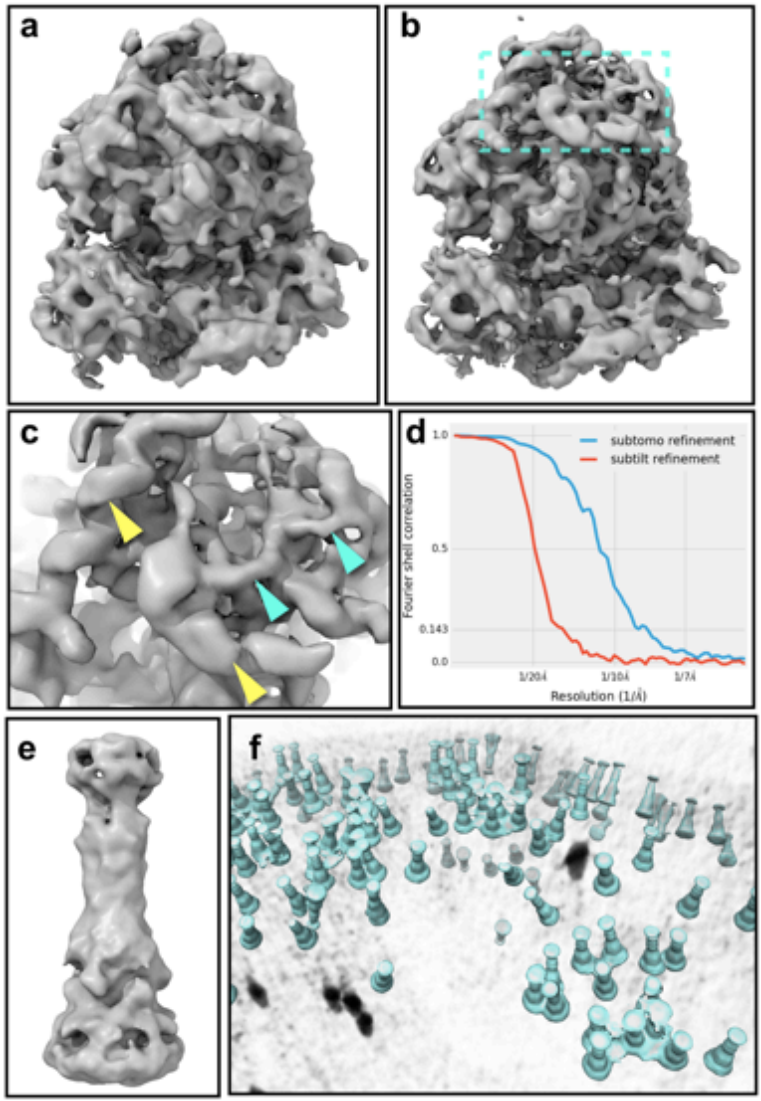

Fig 4. Subtomogram refinement. (a) Subtomogram averaging of ribosome (EMPIAR-10064) before subtilt refinement. (b) Subtomogram averaging after subtilt refinement. (c) Zoomed-in view of (b) with yellow arrows pointing to RNA helices and cyan arrows pointing to resolved alpha-helices. (d) Gold-standard FSC curves of the ribosome subtomogram averaging before (red) and after (blue) subtilt refinement. (e) Subtomogram averaging of the tolc-acrAB drug pump. (f) Location and orientation of the drug pump particles mapped back to a tomogram.

**Methods**

**Tomogram reconstruction**

To seed the iterative tilt-series alignment, a coarse alignment is first performed. First, the unaligned tilt series is downsampled to 512 x 512 pixels, subject to a real-space ramp filter, Fourier bandpass filter, and normalized to mean value of zero and standard deviation of one. A coarse alignment is then performed under a soft Gaussian mask. The alignment begins with the center tilt image (typically near untitled) and propagates sequentially in both directions. After the coarse translational alignment, common lines are used to identify the tilt axis direction. Only angles 0-180 degrees are permitted in this process to ensure no handedness flips occur. Although the handedness is consistent throughout the dataset, it is not necessarily correct due to the 180-degree ambiguity in the tilt axis direction. If the correct orientation of the tilt axis in the images has already been determined for the microscope, it can be specified instead of performing the common-lines search. Finally, the tilt series is reconstructed to produce the preliminary tomogram. The 512 x 512 box size is small enough, that direct Fourier inversion can be used without tiling. Since higher tilt images include information outside the frame of the zero tilt image, higher tilts have a proportional soft mask is applied at the edges of each image parallel to the tilt axis just before reconstruction.

After the initial tomogram reconstruction, an iterative alignment-reconstruction process is performed beginning with 512 x 512 images gradually decreasing downsampling until the fully sampled images are being used (typically 4k x 4k). Each iteration begins with landmark selection in the tomogram from the previous iteration, followed by multiple rounds of landmark location refinement and tilt parameter refinement as described above, and ends with the final downsampled tomogram reconstruction along with the optimized alignment parameters. By default, we perform 2 iterations at 512 x 512, and 1 iteration at 1024 x 1024 , 2048 x 2048 and 4096 x 4096. When the input tilt series is larger than 4096 x 4096, such as DE-64 or K2 super-resolution images, we only perform alignments from 512x512 to 4096 x 4096. It is worth noting that in all iterations, reconstruction of the full tomogram is always done using the pre-filtered 512 x 512 tilt series. These tomograms are only used for selection of landmarks, whose locations are later refined in subtomograms using the appropriate downsampling.

To select landmarks, the 512 x 512 x 256 tomogram is further binned by 4 by taking the minimal value of each 4 x 4 x 4 cube and the result is highpass filtered. In this stage of processing, it is important to note that higher densities have lower values in raw tomograms, which is opposite from the normal EMAN2 convention. Voxel values in the tomogram are sorted and the program picks voxels separated by a minimal distance as landmarks. By default, 20 landmarks are selected and the distance threshold is 1/8 of the longest axis of the tomogram.

Multiple rounds of landmark location refinement and tilt parameter refinement are performed after landmark selection. In each round, we refine the 3D location of landmarks and one of the alignment parameters, including translation, tilt axis rotation, tilt angle and off-axis tilt. Because there is different uncertainty in the determination of each parameter, we begin with refinements tilt image translation and global tilt axis rotation, then refine on and off-axis tilt angles.

In landmark location refinement, we first extract subtilt series of the landmarks from the tilt series and reconstruct the landmarks at the current level of binning. By default, we use box size of 32 for bin-by-8 and bin-by-4 tilt series, 1.5x box size for bin-by-2 and 2x box size for unbinned iterations. We locate the center of landmarks by the coordinate of the voxel with minimal value for bin-by-8 and bin-by-4 iterations and by the center of mass for bin-by-2 and unbinned iterations. This use of center-of-mass rather than aligning features within each landmark region might seem that it could reduce alignment accuracy. However, a common problem with tomographic alignments is that it is possible to have self-consistent alignments with an incorrect translation orthogonal to the tilt axis, producing distorted features in reconstructions when viewed along the tilt axis. Using of center-of-mass for alignment seems to largely avoid this problem, particularly when combined with exclusion of landmarks which are outliers in the alignment process.

To refine the alignment parameters, we first project landmark coordinates to each tilt using currently determined alignment, and extract 2D particles of the same box size at current binning. The center of each 2D particle is determined in the same way that 3D landmarks are centered, and the distance from the center of the 2D particle to the projection of 3D coordinates is computed. For each tilt, the Powell optimizer from Scipy is used to refine alignment parameters and minimize the averaged distance from all landmarks. By default, 10% landmarks with the highest averaged distance in each tilt are ignored during the optimization. The averaged error per tilt is also used in the following round of landmark location refinement and tomogram reconstruction where 10% of tilt images with highest error are excluded.

After all the refinement iterations are finished, the final tomogram is reconstructed. When reconstructing the tomogram by tiling, we use a tile length of 1/4 the tomogram length and pad the 3D cube by an extra 40% during reconstruction. The step size between the tiles is 1/8 tomogram length, and overlapping tiles are shifted by half tile in x and y. 2D tiles are subjected to an edge decay mask along x-axis like the mask used in the full tomogram reconstruction. After reconstruction of each tile, a mask with Gaussian falloff is applied to subvolumes before they are inserted into the final reconstruction. The mask is described by
$$f = 1 + e^{-10(x^2+y^2)} - e^{-10((abs(x)-0.5)^2+(abs(y)-0.5)^2)},$$
where $x, y$ are the coordinate of the voxel from the center of tile, ranging from -1 to 1. This specific shape of mask is used so the summed weight in each voxel in the tomogram is 1, and the soft Gaussian falloff reduces the edge artifacts from the reconstruction of each tile. After reconstruction, the tiles are clipped and added to the final volume to produce the final tomogram. This entire process requires on the order of 10 minutes per tomogram (Table S1).

**Initial model generation for subtomogram averaging**
In the stochastic gradient descent based initial model generation process, we use a very small batch size (12 particles per batch by default) and a learning rate of 0.1 to introduce enough fluctuations into the system. The list of input particles is shuffled before grouping into batches. Particles may be optionally downsampled and lowpass filtered before alignment. Particles in the first batch are averaged in random orientations to produce a map which is then filtered to 100Å and used as the initial alignment reference, which will have roughly the correct radial density profile, but meaningless azimuthal information. In each subsequent batch, particles are aligned to the reference and an average is generated. Any empty regions remaining in Fourier space is filled with information from corresponding Fourier regions in the current reference. We calculate the per voxel difference between the reference and the new averaged map and update the reference toward the average by the learning rate. The program goes through only 10 batches in each iteration by default, except the number of batches is doubled in the first iteration. The first iteration is longer because when symmetry is specified, the program aligns the reference to the symmetry axis after each iteration, and it is necessary to have a map with correct low resolution features to perform a symmetry search stably.

**Subtilt refinement**

The first step of subtilt refinement is to compute the orientation of each subtilt using the orientation of the subtomogram and the alignment of tilt images in the tomogram. The refinement starts from 32 randomly distributed orientations centered around the previous orientation. One of the initial positions is always the previously determined orientation so the worst-case answer is no change. From these positions, an iterative search is performed starting from Fourier box size 64 to full box size, similar to the subtomogram refinement. During the refinement, the reference map is projected using Fourier space slicing with Gaussian interpolation. The comparison between the projection and the 2D particle is scored with CTF weighted Fourier ring correlation for comparison.

We refine even/odd particle sets independently in the subtilt refinement. By default, the program uses all tilt images and removes the 50% of particles with the worst score, generally correlating with tilt angles. There is also an option provided to explicitly exclude high angle tilt images. We also remove subtilt particles with scores beyond 2-sigma around the mean, because practically, particles with very high alignment scores often contain high contrast objects such as gold fiducials, and low score particles are often at the edge of the micrograph and has little signal. Before inserting the images to the 3D Fourier volume, we normalize their scores to (0,1) and weight the particles by their scores when reconstructing the 3D average. The 3D volume is padded by 2 to avoid edge artifacts, and reconstruction is performed with Gaussian interpolation with variable width with respect of Fourier radii. The averaged map is filtered by the "gold-standard" FSC.

**Processing of example data sets**
We processed the 4 "mixedCTEM" from the EMPIAR-10064 purified ribosome dataset. The tomograms were reconstructed from the tilt series automatically using default parameters. 3239 particles were selected via template matching followed by manual bad-particle removal. Defocus values were calculated using default options and the resulting defocus values range from 2.4 to 3.7 µm. CTF-corrected subtomograms were generated with a box size of 180. An initial model was produced using all particles as input, with 3x downsampling and a target resolution of 50Å. Next, 4 rounds of subtomogram refinement and 3 rounds of subtilt refinement were performed to arrive at the final map, which was sharpened using a 1-D structure factor calculated from EMD-5592, masked via EMAN2 auto-masking, and filtered by the local gold-standard FSC.

Tomograms of the AcrAB-TolC pump in E. coli cells were collected on a JEOL3200 equipped with a Gatan K2 camera. Tomogram reconstruction and CTF determination were performed in EMAN2 using default parameters. The unbinned particle data had an Å/pix of 3.365, and a box size of 140 was used during particle extraction. 25 high SNR particles were used for initial model generation. For structures with symmetry, applying the symmetry before the initial model generation converges tends to trap the SGD in a local minimum and not achieve the optimal result. So here a two-step approach was used to build the initial model. First 5 iterations of our SGD routine were performed imposing C1 symmetry. After aligning the result to the symmetry axis, we performed 5 more iterations with C3 symmetry. Subtomogram averaging was then performed using 1321 particles from 9 tomograms while applying c3 symmetry. To focus on the protein while preserving information from the membrane for improved alignment, a mask with values ranging from 0.5-1 around the pump and 0-0.5 covering a larger cylinder was applied to the map each iteration before alignment. The final map was filtered by local FSC and sharpened using a 1-D structure factor obtained from a high-resolution single-particle structure of the purified AcrAB-TolC complex.

Rendering of density maps is performed with UCSF ChimeraX.

**Supplementary Figure**

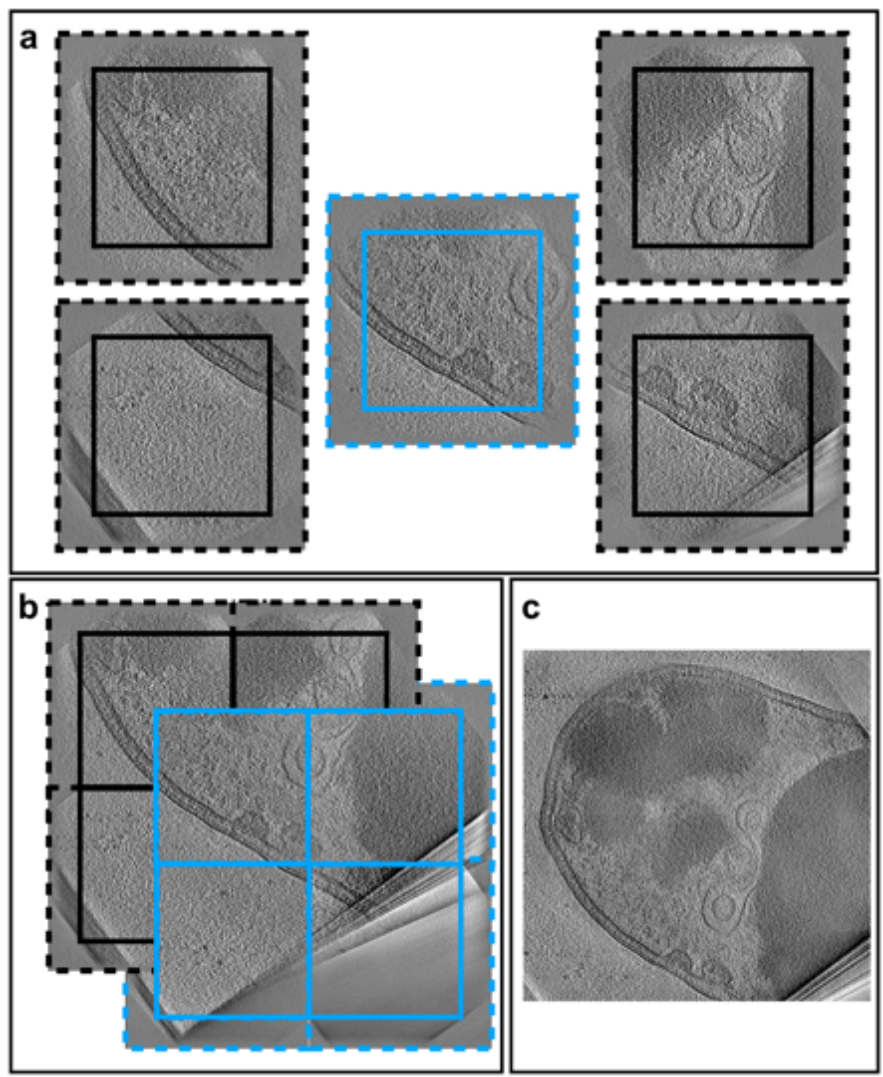

Fig S1. Tiling strategy for tomogram reconstruction. (a) Reconstruction of individual tiles. Each tile is padded to the size of the dashed box during the reconstruction, and clipped to the size of the solid box. (b) Overlapping tiles to reduce edge effects. (c) Resulting by-tile reconstruction.

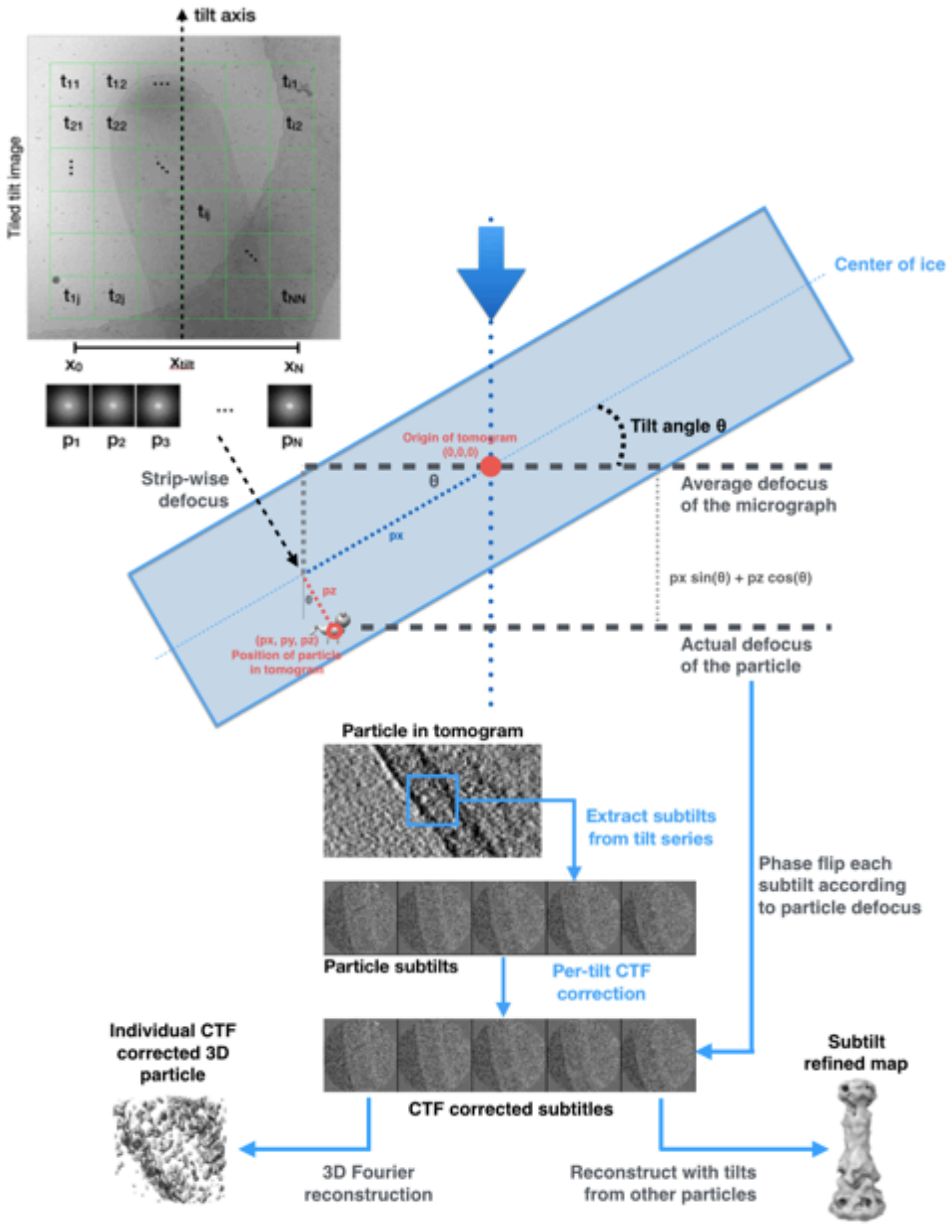

**Fig S2. Subtilt CTF determination.** We measure CTF in each tilt image by tiling the tilt images and calculating coherent power spectra along strips parallel to the tilt axis. These power spectra, geometric information from the tilt angle, and the 3D position of each extracted particle are used to determine per-particle defoci. Once CTF curves have been fit to the data, the parameters are used to phase flip individual particle subtilt images for subsequent processing.

**Table S1**

| Task | Program name | # Cores | Walltime (min) | Iterations |
|---|---|---|---|---|
| **Raw data import** | e2import.py | | 1 | |
| **Tomographic reconstruction** | e2tomogram.py [†] | 12 | 9 | 2,1,1,1 |
| **Reference-based particle picking** | e2spt_tempmatch.py | | 7 | |
| **CTF correction** | e2spt_tomoctf.py | | 2 | |
| **Subtomogram extraction** | e2spt_extract.py [†] | 1 | 31 | |
| **Initial model generation** | e2spt_sgd.py [†] | 12 | 41 | 3 |
| **Subtomogram refinement** | e2spt_refine.py [†] | 12 | 181 | 3 |
| **Subtilt refinement** | e2spt_tiltrefine.py [†*] | 96 | 308 | 6 |

Table S1. Program runtimes. Parallelism: * = MPI, [†] = Thread. Note, e2spt_sgd.py is parallelized by batch, so running the program with a batch size of 12 will use 12 threads.